\documentclass[11pt]{article}

\usepackage[utf8x]{inputenc}
\usepackage[T1]{fontenc}

\usepackage{amsmath, amsfonts, amssymb, amsthm}
\usepackage{mathrsfs}

\usepackage{bbm}
\usepackage{latexsym}
\usepackage{textcomp}
\usepackage{mathabx}%for special needs of set operations - should be reviewed!!!

\usepackage{tensor}

\usepackage{cite}

\usepackage[a4paper, textwidth=17cm,textheight=22cm]{geometry}

\usepackage[affil-it]{authblk}

\DeclareMathOperator{\diag}{diag}

\DeclareMathOperator{\pa}{\partial}

\DeclareMathOperator{\supp}{supp}

\newcommand{\eff}{\mathrm{eff}}
\newcommand{\VEV}[1]{\left\langle {#1} \right\rangle}

\newcommand{\oo}{\circ}
\newcommand{\OO}{\bullet}

\newcommand{\e}{\varepsilon}
\newcommand{\m}{\mu}
\newcommand{\n}{\nu}
\newcommand{\al}{\alpha}

\newcommand{\de}{\delta}
\newcommand{\De}{\Delta}
\newcommand{\et}{\eta}

\newcommand{\g}{\gamma}
\newcommand{\G}{\Gamma}

\newcommand{\la}{\lambda}

\newcommand{\ch}{\chi}
\newcommand{\x}{\xi}

\newcommand{\si}{\sigma}

\newcommand{\te}{\theta}

\newcommand{\z}{\zeta}
\newcommand{\ta}{\tau}

\newcommand{\CD}{\mathcal{D}}

\newcommand{\CT}{\mathcal{T}}

\newcommand{\CO}{\mathcal{O}}

\newcommand{\CA}{\mathcal{A}}

\newcommand{\uvd}{\underline{\mathrm{deg}}}
\newcommand{\ird}{\ol{\mathrm{deg}}}

\newcommand{\bydef}{\stackrel{.}{=}}
\newcommand{\ol}{\overline}
\newcommand{\ul}{\underline}

\newcommand{\lineco}{\boxbackslash}

\newcommand{\IR}{\mathbb{R}}
\newcommand{\IC}{\mathbb{C}}

\newcommand{\II}{\mathbbm{1}}
\newcommand{\IM}{\mathbb{M}}
\newcommand{\IP}{\mathbb{P}}

\newcommand{\IN}{\mathbb{N}}

\newcommand{\SF}{\mathscr{F}}

\newcommand{\SCA}{\mathscr{A}}

\newcommand{\SCL}{\mathscr{C}}

\newcommand{\ST}{\mathscr{T}}
\newcommand{\SG}{\mathscr{G}}

\newcommand{\SL}{\mathscr{L}}
\newcommand{\SZ}{\mathscr{Z}}

\theoremstyle{plain}

\theoremstyle{definition}

\theoremstyle{remark}

\title{BPHZ Renormalization in Configuration Space for the $\CA^4$-Model}
\author[1,2]{Steffen Pottel \thanks{Electronic address: steffen. pottel 'at' itp. uni-leipzig. de}}
\affil[1]{\small Institut f\"ur Theoretische Physik, Universit\"at Leipzig, Postfach 100920, D-04009 Leipzig, Germany }
\affil[2]{\small Max-Planck-Institute for Mathematics in the Sciences, Inselstra\ss e 22, D-04103 Leipzig, Germany}
\date{\today}

\begin{document}

\maketitle

\abstract{\begin{center}\begin{minipage}{0.8\textwidth}
	\noindent
	Recent developments for BPHZ renormalization performed in configuration space are reviewed and applied to the model of a scalar quantum field with quartic self-interaction. 
	An extension of the results regarding the short-distance expansion and the Zimmermann identity is shown for a normal product, which is quadratic in the field operator. 
	The realization of the equation of motion is computed for the interacting field and the relation to parametric differential equations is indicated.
	\end{minipage}\end{center}
}

% change in layout
\setlength{\parindent}{0em}
\setlength{\parskip}{1.0ex plus 0.5ex minus 0.2ex}

\section{Introduction}
\label{se:Intro}

Interacting quantum field theories can be approximated about the free theory with sufficient accuracy in most models.
On Minkowski space, it is convenient to work in momentum space for theories at zero temperature due to the translation invariance.
However, it can be more instructive to work directly in configuration space in situations with different analyticity of the correlation functions, for instance at finite temperature \cite{Collins:1984xc}, or in situation with nontrivial geometries \cite{Brunetti:1999jn}.
Independently of momentum space or configuration space formulation, almost all physical quantities require non-naive formulations in comparison with their possible classical counterparts and employ as a rule regularization methods.
A renormalization scheme guarantees that the latter are compatible with the properties of a considered model so that the physical quantities are rendered well-defined.
The method developed by Bogoliubov, Parasiuk, Hepp and Zimmermann (BPHZ) \cite{Bogoliubov:1957gp,Hepp:1966eg,Zimmermann:1968mu,Zimmermann:1969jj} does not only have a well accessible regularization technique but also resolves the combinatorial problem behind occurring singularities in weighted Feynman graphs by the forest formula.
Both features turned out to be very useful in the study of vector-type gauge theories \cite{Becchi:1975nq,Tyutin:1975qk} and anomalies \cite{Piguet:1980nr} as well as the definition of composite operators \cite{Zimmermann:1972te} and establishing a relation to Wilson's operator product expansion \cite{Wilson:1972ee,Zimmermann:1972tv}.
Recently, the ideas of BPHZ renormalization have been translated into a configuration space formulation \cite{Pottel:2017aa,Pottel:2017bb,Pottel:2017cc}, which avoids introducing an auxiliary mass term for theories including massless fields \cite{Lowenstein:1975rg,Lowenstein:1975ps,Lowenstein:1975ku}. 
It is the aim of this work, to present the construction of configuration space BPHZ renormalization at the example of a scalar field theory with quartic self-interaction.
In particular, this approach admits a detailed discussion of counterterms and the derivation of composite operators.  

The paper is organized as follows. 
In the next Section, we motivate the subsequent results by a discussion of the perturbative argument applied to the nonlinear equation of motion on the classical level followed by a formulation of the problem set for renormalization in configuration space. 
Afterwards, we review the BPHZ-method in configuration space in the Section \ref{se:BPHZ} and discuss the ambiguities in Section \ref{se:Ambi}.
We derive the results on the short-distance expansion and the Zimmermann identity for a special case in the considered model in the fifth and sixth Section, respectively.
Finally, we return to the equation of motion in Section \ref{se:Conc} and conclude with a discussion on differential operations.

\section{Preliminaries}
\label{se:Pres}

Consider a real, classical scalar field $\CA$ with mass $m^2\geq 0$ on four-dimensional Minkowski space $(\IM,\et)$ with $\IM=\IR^4$ and $\et=\diag(1,-1,-1,-1)$.
We assign the Lagrangian density
\begin{align}\label{eq:ClassLagrangian}
	\SL[\CA](x) \bydef \underbrace{\frac{1}{2} \pa^\mu \CA(x) \pa_\mu \CA(x) - \frac{1}{2} m^2 \CA^2(x)}_{\bydef \SL_0} \underbrace{- \frac{\la}{4!} \CA^4(x)}_{\bydef \SL_I}
\end{align}
to the interacting field $\CA$ and distinguish between the \emph{free} part $\SL_0$ and the \emph{interaction} $\SL_I$.
The equation of motion of the classical field is then given by
\begin{align}\label{eq:ClassicalNonLinEoM}
	(\Box + m^2) \CA(x) = \frac{\la}{3!} \CA^3(x),	\quad \Box \bydef \et^{\m\n} \pa_\m \pa_\n,
\end{align}
where a well-posedness theory for global solutions is generally not available. 
Therefore we approximate solutions of \eqref{eq:ClassicalNonLinEoM} by a formal expansion about the linear problem
\begin{align}\label{eq:ClassFreeEoM}
	(\Box + m^2) A(x) = 0,
\end{align}
which has smooth global solutions for the Cauchy problem with compactly supported, smooth initial data on a constant time hypersurface \cite[Chapter 3]{Bar:2007zz}.
In particular, the linear problem admits global advanced and retarded fundamental solutions
\begin{align}\label{eq:ClassFundSol}
	(\Box + m^2) G_{A/R}(x) = \de_x,
\end{align}
where $\de_x$ is the Dirac-$\de$-distribution.
With \eqref{eq:ClassFreeEoM} and \eqref{eq:ClassFundSol}, we approximate solutions to \eqref{eq:ClassicalNonLinEoM} in the following way. We expand $\CA$ in a formal power series
\begin{align}\label{eq:PowerSeriesCA}
	\CA = \CA_0 + {\la^1} \CA_1 + {\la^2} \CA_2 + ...
\end{align}
and plug this expansion into \eqref{eq:ClassicalNonLinEoM}. Resolving this order by order, we obtain at zeroth order
\begin{align}\label{eq:ZerothOrder}
	(\Box+m^2) \CA_0 = 0
\end{align}
such that $\CA_0 = A$, thus at first order
\begin{align}\label{eq:FirstOrder}
	(\Box + m^2) \CA_1 = \frac{1}{3!} \CA_0^3 = \frac{1}{3!} A^3.
\end{align}
Restricting $A^3$ to have finite support by multiplication with a test function $f$, solutions to \eqref{eq:FirstOrder} can be constructed by
\begin{align}\label{eq:CAOne}
	\CA_1 = \frac{1}{3!} G_\bullet \ast (fA^3).
\end{align}
The choice of fundamental solution $G_\bullet$ depends on the problem at hand and is specified at a later stage. 
In the same manner one may proceed at second order and constructs solutions $\CA_2$ out of $\CA_1$ and $\CA_0$, thus $G_\bullet$ and $A$. 
We use the knowledge about the order by order solutions for the computation of correlation functions
\begin{align}
	\VEV{\CA(x_1)...\CA(x_n)},
\end{align}
i.e. we expand each field $\CA$ and compute the correlation again order by order. For instance, a contribution to the second order of the two-point function $\VEV{\CA(x)\CA(y)}$ is computed to be
\begin{align}
	\langle \CA_1(x) \CA_1(y) \rangle & = \frac{1}{3!3!} \langle (G_\bullet \ast fA^3)(x) (G_\bullet \ast fA^3)(y) \rangle. 
\end{align}
Spelling out the convolutions, we get 
\begin{multline}
	\left\langle \int_{\IM}\limits G_\bullet(x - z_1) A^3(z_1) f(z_1) dz_1 \int_{\IM}\limits G_\bullet(y - z_2) A^3(z_2) f(z_2) dz_2 \right\rangle \\
	= \int_{\IM}\limits \int_{\IM}\limits \, G_\bullet(x-z_1) G_\bullet(y-z_2) \underbrace{\langle (fA^3)(z_1) (fA^3)(z_2) \rangle}_{=3!3!\langle (A)(z_1) (A)(z_2) \rangle^3  f(z_1) f(z_2) } \, dz_2 dz_1
\end{multline}
so that
\begin{align}
	\langle \CA_1(x) \CA_1(y) \rangle & = \int_{\IM}\limits \int_{\IM}\limits G_\bullet(x-z_1) G_\bullet(y-z_2) G_\bullet^3(z_1-z_2) f(z_1) f(z_2) \, dz_2 dz_1. \label{eq:FirstGraph}
\end{align}
This example of a second order contribution for the two-point function already indicates the problem sets we are going to face throughout this work.
Since the fundamental solution $G_\bullet$ in \eqref{eq:FirstGraph} is defined as a distribution, we have to make sense out of its powers, i.e. products of distributions defined at the same point. 
At the same time, the recursive structure in the definition of $\CA_j$ admits an equivalent description of the combinatorial problem occurring at higher orders, where we identify each fundamental solution $G_\bullet(x_v - x_w)$ in \eqref{eq:FirstGraph} with an edge connecting the vertices $v$ and $w$ in a graph. 

In the next step, we promote the free, classical field $A$ to a distribution writing informally
\begin{align}
	A(f) = \int_{\IM}\limits A(x) f(x) dx
\end{align}
with $f$ being a test function. The quantum field $A(f)$ inherits the following conditions
\begin{align}
	A(f)^* & = A(\ol f), \label{eq:RealityCond} \\
	A((\Box + m^2)f) & = 0, \label{eq:WeakFieldEquation} \\
	A(a f_1 + b f_2) & = a A(f_1) + b A(f_2) \mbox{ with } a,b\in \IC \label{eq:Linearity}
\end{align}
and its quantum character is expressed by
\begin{align}
	[A(f_1),A(f_2)] & = iG(f_1,f_2),\label{eq:FreeFieldCommutator}
\end{align}
where $G$ is the commutator function defined as the difference of advanced and retarded fundamental solution. 
We observed in \eqref{eq:FirstOrder} that the perturbative definition of correlation functions involves powers of the the field $A(f)$, which naively are not well-defined.
Instead consider the two-point function
\begin{align}
	\VEV{A(f)A(g)},
\end{align}
which exhibits singularities if $\supp(f)\cap\supp(g)\neq\emptyset$, such that the Wick ordered product
\begin{align}
	:A(f)A(g): \bydef A(f)A(g) - \VEV{A(f)A(g)}
\end{align}
becomes well-defined in the intersection of the supports and, in particular, for $f=g$. 
In general, the construction of Wick products is recursively defined by
\begin{align}
	:A(f):                       & \bydef A(f)\\
	:A(f_1)...A(f_n): A(f_{n+1}) & = :A(f_1)...A(f_{n+1}):\nonumber\\
	                             & + \sum_{j=1}^n :A(f_1)...\widecheck{A(f_j)}...A(f_n): \VEV{A(f_j) A(f_{n+1})}, 
\end{align}
where $\widecheck\bullet$ denoted the extraction of that field from the Wick polynomial. 
This admits so-called Wick powers $:A^k(f):$ with $k\in\IN$. 
For the construction of an interacting quantum field theory we further need the notion of time-ordered products of quantum fields given by
\begin{align}
	\underbrace{\VEV{\CT A(f)A(g)}}_{\bydef G_c(f,g)} \bydef \te(x^0-y^0) \VEV{A(f)A(g)} + \te(y^0-x^0) \VEV{A(g)A(f)}
\end{align}
and relate to the definition of Wick products by
\begin{align}
	\CT A(f)A(g) = :A(f) A(g): + G_c(f,g)
\end{align}
from which follows that
\begin{align}
	\VEV{:A(f) A(g):} = 0.
\end{align}
All higher orders are then computed recursively.
With the notions defined above, we are able to make sense out of correlation functions of the interacting quantum field $\CA$ in the perturbative approach via the Gell-Mann-Low formula
\begin{align}\label{eq:GellMannLow}
	\VEV{\CT \CA(x_1)...\CA(x_n)} = \frac{\VEV{\CT A(x_1)...A(x_n)\exp\{i\SL_I[A,\la](g)\}}}{\VEV{\CT \exp\{i\SL_I[A,\la](g)\}}}.
\end{align}
Note that $\SL_I[A,\la](g) = \frac{\la}{4!}:A^4(g):$ is defined Wick ordered and is evaluated via the coupling function $g\in\CD(\IM)$.
Then any $n$-point function at fixed order of $\la$ can be written as a sum over products of Feynman propagators $G_c$, which carries exactly the same description in terms of graphs as we found in \eqref{eq:FirstGraph}, i.e.
\begin{align}\label{eq:SpecialWickThm}
	\VEV{\CT A(x_1)...A(x_n) \SL_I[A,\la](x_{n+1})...\SL_I[A,\la](x_{n+m})} = \sum_{\G\in\SG} \prod_{e\in E(\G)} G_c(x_{s(e)},x_{t(e)})
\end{align}
with $\SG$ the set of all graphs $\G$, $E(\G)$ the set of edges in the graph $\G$ and $s,t:E(\G)\rightarrow V(\G)$ boundary maps from the edge set $E(\G)$ to the vertex set $V(\G)$ for an arbitrarily assigned direction to the graph $\G$.
For later purposes, we remark that each graph $\G$ may be composed out of several connected components $\G_1,...,\G_c$, $c\in\IN$.

In order to understand the necessity of regularization and the problem of renormalization in pertubative quantum field theory, we have to discuss the right-hand side of \eqref{eq:SpecialWickThm}.
We indicated already above that the pointwise product of Feynman propagators $G_c$ is not defined naively.
Furthermore, one can show that a single propagator is a well-defined distribution on Minkowski space up to the origin in the sense of boundary values of analytic functions, i.e. introducing an analytic continuation parametrized by $\e$ we find
\begin{align}
	\lim_{\e\rightarrow 0} G_{c,\e} \in \CD'(\IM\setminus \{0\}).
\end{align}
We want to use an argument from microlocal analysis \cite[Chapter 8]{hormander1990analysis}, which turns the product of Feynman propagators into a sensible expression if we find a suitable $\e$-regularization.
Specifically, the argument requires precise knowledge of the wavefront set of $G_c$, but can be simplified to a statement involving only the singular support of $G_c$, i.e. the set of points $x\in\IM$ for which $G_c(x)$ is singular.
It is the idea to choose the $\e$-regularization such that the singular support of the product of Feynman propagators is in the complement of their domain.
Therefore consider the analytic continuation 
\begin{align}
	\et \mapsto \et^\e = \diag(1-i\e,-1,-1,-1)
\end{align}
of the metric, for which one can show \cite{Pottel:2017bb} that for any $x\in\IM$ there exist constants
\begin{align}
	\hat C(\e) & = \left(\frac{1}{\e} + \sqrt{1+\frac{1}{\e^2}}\right)^{-1},\\
	\check C(\e) & = \sqrt{1+\e^2}
\end{align}
such that 
\begin{align}\label{eq:AnaContArgument}
	\hat C(\e) \|x\|^2_\de \leq |z^2| \leq \check C(\e) \|x\|^2_\de
\end{align}
for $z^2 = \et^\e_{\m\n} x^\m x^\n$ and $\|\bullet\|_\de$ denoting the Euclidean norm of a vector in $\IR^4$.
We conclude the argument using the explicit form of the Feynman propagator \cite{Bogolyubov:1980nc}, i.e.
\begin{align}\label{eq:PropagatorExplicit}
	G_c(z) = \frac{m}{4\pi^2 \sqrt{-z^2}} K_1(m\sqrt{-z^2})
\end{align}
with $K_\bullet$ being the modified Bessel function of second kind.
For any real, positive argument $x$, $K_1(x)$ is a strictly decreasing function mapping to the positive real numbers. 
With the help of \eqref{eq:AnaContArgument}, it is then possible to estimate \cite{Pottel:2017bb}
\begin{align}\label{eq:EuclEstBessel}
	K_1(\sqrt{\check C(\e)}\|x\|_\de) \leq |K_1(\sqrt{-z^2})| \leq K_1(\sqrt{\hat C(\e)} \|x\|_\de),
\end{align}
thus find Euclidean bounds on the Feynman propagator, which are singular only at the origin.
Hence the product of Feynman propagators
\begin{align}
	\prod_{e\in E(\G)} G_{c,\e}(x_{s(e)},x_{t(e)})
\end{align}
together with the $\e$-regularization is well-defined everywhere except for configurations with $x_{s(e)}=x_{t(e)}$ for some $e\in E(\G)$.
Let us denote by $\oo$ the set of configuration over $\G$, for which there exists an edge $e\in E(\g)$ such that $x_{s(e)}=x_{t(E)}$, and by $\OO$ the set of configurations over $\G$, for which $x_{s(e)}=x_{t(e)}$ for all $e\in E(\G)$, such that
\begin{align}\label{eq:DomainOfGraphWeight}
	\prod_{e\in E(\G)} G_{c,\e}(x_{s(e)},x_{t(e)}) \in \CD'(\IM^{|V(\G)|}\setminus \oo).
\end{align}
From \eqref{eq:DomainOfGraphWeight}, we read off the problem of renormalization.
Namely, we need to find a prescription for the extension of the distribution to the whole space $\IM^{|V(\G)|}$, which additionally requires a recursive or iterative procedure knowing that we rely on standard results regarding extensions of distributions to a single point \cite[Chapter 3]{hormander1990analysis}, i.e. for distributions $\CD'(\IM^{|V(\g)|}\setminus\OO)$, $\g\subseteq\G$, in our case.
In \cite[Theorem 5.2]{Brunetti:1999jn}, it is stated that there exist a unique extension of a distribution $u_0\in\CD'(\IM^n\setminus\{0\})$ to $u\in\CD'(\IM^n)$ if $u$ has a suitable \emph{scaling degree}, which essentially measures, for a tempered distribution $u_0$, by which inverse polynomial $u_0$ can be bounded in the neighborhood of the point of extension.
Then uniqueness of the extension follows from a comparison of the scaling degree to the dimension of the integration measure, i.e. in order to reach a configuration $\OO$ for a graph $\g\subseteq\G$, all vertices of $\g$ have to have the same loci or, equivalently, all but one vertex of $\g$ have to be integrated such that a configuration $\OO$ is possible. 
The condition for the uniqueness can be specified further in our model, restricting to a graph $\G$ with weight $u_0^\e[\G]$, where $\e$ indicates the analytic continuation.
Since every edge is given by \eqref{eq:PropagatorExplicit}, we know that its scaling degree is $2$, i.e. for small $|x|$, $G_{c,\e}(|x|)$ can be bounded by $|x|^{-2}$.
Further every integration over a single vertex gives dimension $4$ and, for completeness, every derivative acting on a propagator increases the scaling degree by $1$.
Thus we find the \emph{UV-degree of divergence}
\begin{align}
	\uvd(u_0^\e[\G]) = 2 |E(\G)| - 4(|V(\G)|-1) + |\pa|
\end{align}
for the distribution $u_0^\e[\G]\in\CD'(\IM^{|V(\G)|}\setminus \oo)$.
Now suppose that we have $u_0^\e[\g]\in\CD'(\IM^{|V(\g)|}\setminus\OO)$ for some $\g\subseteq\G$.
Then uniqueness of the extension follows from the condition $\uvd(u_0^\e[\g])<0$, which we want to reformulate for the case that $u_0^\e[\g]$ is a tempered distribution.
It follows from classical results in real analysis that a negative UV-degree of divergence is equivalent to local integrability of the distribution kernel so that the extension problem can be restated as a problem of local integrability of the distribution kernel in the considered model. 
Of course, there exist graphs $\g$ for which $u_0^\e[\g]$ has non-negative UV-degree of divergence such that we have to employ a technique which reduces the UV-degree of divergence sufficiently.

\section{The BPHZ-Method in Configuration Space}
\label{se:BPHZ}

In order to find a solution to the extension problem, we have to show local integrability of the distribution kernel $u_0^\e[\G]$ for any integration over a vertex set $I\subset V(\G)$, thus for any sequence
\begin{align}\label{eq:IntegrationSequence}
	\emptyset \subset I_1 \subset ... \subset I_{k-1} \subset V(\G)
\end{align}
with $|V(\G)|=k$. 
In the case of subgraphs $\g\subseteq\G$ with $\uvd(u_0^\e[\g])\geq 0$, we employ a regularization which is based on a variation of the standard Hadamard regularization of singular integrals \cite{Bogoliubov:1957gp}, i.e. we replace the distribution kernel with a suitably chosen Taylor remainder of it but, in particular, not evaluated on test functions in the first place. 
The Taylor operator is defined as
\begin{align}
	t^d_{x|\ol x} \bydef \sum_{|\al|=0}^d \frac{(x-\ol x)^\al}{\al!} D^\al_{x|\ol x},
\end{align}
with $\al$ a multiindex and subtraction point $\ol x$. 
Recall from the momentum space prescription that subtractions are performed at zero external momentum of the considered graph.
Translated into coordinate space, this corresponds to the center of mass of the considered subgraph, which turns into the mean coordinate of the involved vertices for models with a single quantum field \cite[Section 10.3]{Steinmann:2000nr}.
But this choice does not admit non-trivial graph manipulations like fusing or splitting interaction vertices, which we want to be able to perform in view of subsequent applications.
Therefore we introduce the \emph{weighted mean coordinate}
\begin{align}\label{eq:SubtractionPoint}
	\ol x_\g \bydef \frac{1}{2|E(\g)|} \sum_{v\in V(\g)} |E(\g|v)| x_v
\end{align}
for a graph $\g\subseteq\G$, where $E(\g|v)$ denotes the set of edges in $\g$ incident to the vertex $v\in V(\g)$. 
We observe that the subtraction point $\ol x_\g$ is chosen to coincide with the singular configuration of $u_0^\e[\g]$ so that a direct application of the Taylor operator to the distribution kernel is not defined.
However, the distribution kernel of the edge complement $u_0^\e[\G\lineco\g]$ ($\lineco$ denoting the set difference with respect to the sets of edges) is by construction smooth at the singular point of $u_0^\e[\g]$.
Hence we modify the Taylor operation by the operator $\IP(\bullet)$ taking an element $\g\subseteq\G$ and mapping to the edge complement $\G\lineco\g$, i.e. 
\begin{align}
	t(\g) \bydef t^{d(\g)}_{\g|\ol \g} \IP(\g)
\end{align}
with $d(\g) \bydef \lfloor \uvd(u_0^\e[\g]) \rfloor$ so that
\begin{align}
	t(\g) u_0^\e[\G] = u_0^\e[\g] t^{d(\g)}_{\g|\ol\g} u_0^\e[\G\lineco\g].
\end{align} 
We call $\g\subseteq\G$ a \emph{renormalization part} if $d(\g)\geq 0$ and otherwise we set $t(\g)=0$.
In contrast to the momentum space scheme, the subtraction point $\ol x_\g$ depends still on the loci of all vertices in $V(\g)$, which results in a different set of constraints on the composition of Taylor operations.
Namely, we define overlap of renormalization parts with respect to vertices instead of edges, i.e. two renormalization parts $\g$ and $\g'$ are \emph{overlapping} if none of the following conditions hold
\begin{align}
	V(\g) \subset V(\g'), \quad V(\g) \supset V(\g'), \quad V(\g)\cap V(\g') = \emptyset.
\end{align}
It is worth noting that renormalization parts are determined by the set of vertices and all edges connecting those, so-called \emph{full vertex parts}.
With this, we are in the position to rewrite Bogoliubov's $R$-operation in the spirit of Zimmermann \cite{Zimmermann:1969jj} by the forest formula with notions in configuration space
\begin{align}\label{eq:ForestFormula}
	Ru_0^\e[\G] \bydef \sum_{F\in\SF} \prod_{\g\in F} (-t(\g)) u_0^\e[\G],
\end{align}
where $\SF$ is the set of all forests $F$, which are sets of non-overlapping renormalization parts.
In order to show local integrability with respect to any $I\subset V(\G)$, \eqref{eq:ForestFormula} has to be reordered such that graphs near singular configurations are computed as the Taylor remainder instead of the Taylor polynomial. 
We want to formalize this statement in the following.
A vertex $v\in V(\G)$ is called integrated if $v\in I$ and constant if $v\in V(\G)\setminus I$.
With this, we call a graph $\g\subseteq \G$ \emph{integrated} if all vertices of $\g$ are integrated, \emph{variable} if all but one vertices are integrated and \emph{constant} otherwise.
In particular, we are interested in \emph{maximal variable} graphs, i.e. a variable graph $\g$ such that there exists no variable graph $\g'$ with $\g\subset \g'$.
Then one can show \cite{Pottel:2017aa} that the forest formula \eqref{eq:ForestFormula} can be rewritten as
\begin{align}\label{eq:SaturatedForestFormula}
	Ru_0^\e[\G] = \sum_{F'\in \SF'} \prod_{\g\in F'} \ch(\g) u_0^\e[\G],
\end{align}
where
\begin{align}
	\ch(\g) = 
	\begin{cases}
		1-t(\g) & \mbox{for $\g$ maximal variable} \\
		-t(\g) & \mbox{otherwise}
	\end{cases}
\end{align}	
The set of maximal variable renormalization parts depends on $I$ and it is important to note that the set is not unique such that \eqref{eq:SaturatedForestFormula} holds only in a small neighborhood in configuration space about a set of maximally variable, mutually disjoint renormalization parts.
This new form bears the advantage that for any $I\subset V(\G)$
\begin{align}\label{eq:ChiOfGraph}
	\uvd_I(\ch(\g) u_0^\e[\G]) \leq
	\begin{cases}
		\uvd_I(u_0^\e[\g]) - d(\g) - 1 & \mbox{for $\g$ maximal variable} \\
		\uvd_I(u_0^\e[\g]) & \mbox{otherwise},
	\end{cases}
\end{align}
where $\uvd_I$ indicates the scaling only with respect to vertices in $I$.
The relation \eqref{eq:ChiOfGraph} holds recursively throughout the products in each summand of the modified forest formula \eqref{eq:SaturatedForestFormula} so that
\begin{align}\label{eq:NegativeDegDivergence}
	\uvd_I(Ru_0^\e[\G]) < 0
\end{align} 
holds for each aforementioned small neighborhood.
But it holds in particular in the complement of these small neighborhoods since those do not contain any maximal variable renormalization parts such that we obtain \eqref{eq:NegativeDegDivergence} on the whole space.
Then local integrability of $Ru_0^\e[\G]$ follows from the observation that \eqref{eq:NegativeDegDivergence} holds for each element in a sequence \eqref{eq:IntegrationSequence} and the extension to the whole space $Ru_0^\e[\G]\mapsto Ru^\e[\G]\in \CD'(\IM^{|V(\G)|})$ as well as 
\begin{align}
	\lim_{\e\rightarrow 0} Ru^\e[\G] \in \CD'(\IM^{|V(\G)|})
\end{align}
follow from standard arguments of distribution theory \cite{Pottel:2017bb}. Transferring this result to the $A^4$-model, we find that
\begin{align}
	\lim_{\e\rightarrow 0} R\prod_{e\in E(\G)}G_{c,\e}(x_{s(e)},x_{t(e)}) \in \CD'(\IM^{|V(\G)|})
\end{align}
so that, summing over all graphs $\G$ in \eqref{eq:SpecialWickThm}, 
\begin{align}\label{eq:RenormalizedVEV}
 	\VEV{\CT_R A(f_1)...A(f_n) \SL_I[A,\la](g)...\SL_I[A,\la](g)}
\end{align} 
is a well-defined expression for all test functions $f_j$ and $g$.
We notice that the construction does not distinguish between positive and vanishing mass parameter $m$, since the behavior for small distances among vertices is the same, and that the interaction is still restricted to the support of the coupling function $g$. 
In the remainder of this section, we discuss the possibility of the constant coupling limit, i.e. $g\rightarrow 1$ in \eqref{eq:RenormalizedVEV}.
Again, it follows from classical results in real analysis that the limit is well-defined if the distribution kernel, expanded in a sum over graphs, is absolutely integrable.
In analogy to the treatment of small momenta \cite{Lowenstein:1975ps}, one can show \cite{Pottel:2017aa} that the renormalized distribution kernel $Ru_0^\e[\G]$ is absolutely integrable if the unrenormalized kernel $u_0^\e[\G]$ is sufficiently fast decaying for long ranges, i.e. denoting the long range decay behavior by $\ird(\bullet)$, one obtains
\begin{align}
	\ird(Ru_0^\e[\G]) \geq \ird(u_0^\e[\G]),
\end{align}
where $\uvd(u)<0$ and $\ird(u)>0$ are the sufficient conditions for integrability. 
Returning to the $A^4$-model, this translates into the problem whether 
\begin{align}
	\prod_{e\in E(\G)} G_{c,\e}(x_{s(e)},x_{t(e)})
\end{align}
can be controlled for large arguments $x_{t(e)}-x_{s(e)}$. 
In the previous section we found Euclidean bounds for the Feynman propagator using \eqref{eq:EuclEstBessel}, which tells us that the propagator is exponentially decaying for large arguments if the mass $m$ is positive so that $Ru_0^\e[\G]$ is absolutely integrable over the set of interaction vertices $\SL_I[A,\la]$ and
\begin{multline}\label{eq:ExplicitCouplingLimit}
	\lim_{g\rightarrow 0} \VEV{\CT_R A(f_1)...A(f_n) \SL_I[A,\la](g)...\SL_I[A,\la](g)} \\
	= \VEV{ \CT_R A(f_1)...A(f_n) \int \SL_I[A,\la](y_1) dy_1 ...\int \SL_I[A,\la](y_m) dy_m }.
\end{multline}
However, this exponential decay does not survive in the limit when the mass parameter $m$ tends to zero.
While 
\begin{align}
	\lim_{m\rightarrow 0} G_{c,\e}(m\sqrt{-z^2}) = -\frac{1}{4\pi^2 z^2 }
\end{align}
continuously, we notice that it can be bounded by $|x^2|^{-1}$ after the limit.
Therefore the constant coupling limit exists even in the massless case, but is restricted to graphs which do not contain bilinear, derivative-free (interaction) vertices $:A^2(g):$.

\section{Discussion of Ambiguities}
\label{se:Ambi}

The BPHZ method in configuration space admits a unique extension of $Ru_0^\e[\G]$ but not of $u_0^\e[\G]$ in general.
Indeed, we observe that we may modify $u_0^\e[\g]$ by adding a term
\begin{align}\label{eq:NonUniquenessDeltaTerm}
	u_0^\e[\ol \g] = \sum_{|\al|=0}^{d(\g)} c_\al D^\al \de_{\ol x_\g}
\end{align}
for a renormalization part $\g\subseteq\G$, since 
\begin{align}
	(1-t(\g))(u_0^\e[\G\lineco\ol \g]u_0^\e[\ol\g]) = 0.
\end{align}
The constants $c_\al$ are fixed after employing suitable normalization conditions, which are discussed below.
The drawback of writing the freedom in form of \eqref{eq:NonUniquenessDeltaTerm} is the lack of interpretation in terms of fields.
Therefore we rewrite 
\begin{align}\label{eq:NonuniquenessFieldTerm}
	u_0^\e[\ol \g] = \sum_{|\al|=0}^{d(\g)} c_\al D^\al :A^{|\ol\SCA(\g)|}(\ol x_\g):,
\end{align}
where $\ol\SCA(\g)$ is the set of fields $A$ corresponding to both vertices in $V(\g)$ and external lines of the full vertex part $\g$. 
It follows from the definition of the $R$-operation that \eqref{eq:NonuniquenessFieldTerm} is a sensible expression, i.e. the Taylor operators are defined in such a way that external lines of a renormalization part $\g\subseteq\G$, which are in the edge set complement $\G\lineco\g$, get fused to a new vertex such that
\begin{align}\label{eq:GraphReduction}
	\G\lineco\g \mapsto \G/\g,
\end{align}
where $\G/\g$ denotes the graphs $\G$ with $\g$ contracted to a point.
We may additionally rule out some of the terms in \eqref{eq:NonuniquenessFieldTerm}, which are not compatible with the $A^4$-model on Minkowski space, and proceed by the number of external edges of a renormalization part.
In this respect, the UV-degree of divergence for $\g\subset\G$ becomes
\begin{align}\label{eq:UVDegreeVertex}
	\uvd(u_0^\e[\g]) = 4 + \sum_{v\in V(\g)} (\dim(v) - 4) - \ol d_\g,
\end{align}
where $\dim(v)$ is given by the number of fields $A$ and the number of derivatives $\pa$ at the vertex $v$, and $\ol d_\g$ is the \emph{codegree}, determined by the number of elements in $\ol\SCA(\g)$ and the number of derivatives acting on elements in $\ol\SCA(\g)$.
We remark that the $R$-operation still solves the extension problem if we assign subtraction degrees
\begin{align}
 	\de(\g) > d(\g)	
\end{align} 
to renormalization parts $\g$ as long as 
\begin{align}
	\de(\g) \geq d(\g/\g_1...\g_c) + \sum_{j=1}^c \de(\g_j)
\end{align}
holds in each forest for maximal subgraphs $\g_j\subset\g$.
Keeping this fact in mind, we simplify \eqref{eq:UVDegreeVertex} to
\begin{align}\label{eq:UVDegreeReduction}
	\de(\g) = 4 - 3 |V_\mathrm{ext}(\g)| - \ol d_\g
\end{align}
by assigning the subtraction degree $4$ to every vertex in $\g$ with valency greater than $1$ and $V_\mathrm{ext}$ denoting the set of external vertices in $\g$, i.e. vertices with valency $1$.
With this, it is reasonable to discuss admissible ambiguities of $u_0^\e[\G]$ in form of \eqref{eq:NonuniquenessFieldTerm} order by order.
Due to \eqref{eq:GellMannLow}, we start at first order, where we consider a graph with two external vertices. 
This admits a renormalization part $\g$ with subtraction degree $\de(\g)=0$ and codegree $\ol d_\g=1$, which is the only type that admits an external vertex in the renormalization part in the model.  
The corresponding ambiguity is given by
\begin{align}
	\int c_1 A(\ol x_\g) d\ol x_\g
\end{align}
and constitutes a correction to the external fields, which vanishes after employing suitable normalization conditions.
For codegree $\ol d_\g=2$, we obtain subtraction degree $2$ and
\begin{align}
	\int (c_2 :A^2(\ol x_\g): + c_3 :\pa^\m A(\ol x_\g) \pa_\m A(\ol x_\g): ) d\ol x_\g,
\end{align}
where $A\pa A$ is ruled out by Lorentz invariance and $A\Box A$ is related to $\pa A\pa A$ via integration by parts.
Finally, we find for codegree $\ol d_\g=4$ the ambiguity
\begin{align}
	\int c_4 :A^4(\ol x_\g): d\ol x_\g,
\end{align}
since a renormalization part with codegree $3$ does not exist.
Note that these ambiguities recur at increasing orders of the coupling $\la$.
Therefore we extend the interaction Lagrangian $\SL_I[A,\la]$ to the effective Lagrangian 
\begin{align}\label{eq:EffectiveLagrange}
	\SL_{\eff}[A,\la](x) = \frac{1}{2} a N_4[\pa^\m A(x) \pa_\m A(x)] - \frac{1}{2} b N_4[A^2(x)] - \frac{1}{4!} c N_4[A^4(x)],
\end{align}
where $a,b = \CO(\la)$, $c = \la + \CO(\la^2)$ and we adopted the notion of normal products $N_\de[\bullet]$ \cite{Lowenstein:1971vf}.
Then all graphs of the renormalized $A^4$-model are covered by (omitting test functions at the external fields)
\begin{align}\label{eq:EffectiveGellMannLow}
 	\VEV{\CT_R \CA(x_1)...\CA(x_n)} = \frac{\VEV{\CT_R A(x_1)...A(x_n) \exp\left\{ i\int \SL_\eff[A,\la](y)dy \right\}}}{\VEV{\CT_R \exp\left\{ i\int \SL_\eff[A,\la](y)dy \right\}}}
\end{align} 
including the ambiguities, which stem from the definition of the $R$-operation.
The ambiguities are fixed by the following normalization conditions.
At the normalization scale $\m^2>0$, we demand for the 2-point function that
\begin{align}
	(\Box + \m_\mathrm{phys}^2) G^{(2)}(x,y) = (\m^2-m_\mathrm{phys}^2) G^{(2)}(x,y) \quad \mbox{for } x\neq y
\end{align}
such that additionally for $\m^2\rightarrow m^2_\mathrm{phys}$
\begin{align}
	(\Box + m^2_\mathrm{phys}) G^{(2)}(x,y) = -i \de(x,y).
\end{align}
Finally, the normalization condition for the 4-point function reads
\begin{align}
	G^{(4)}(x_1,x_2,x_3,x_4)_{|x_j=\ol x} = i\la.
\end{align}

\section{Short-Distance Expansion}
\label{se:OPE}

We want to discuss the possibility of defining interacting quantum fields $\CA$ at the same spacetime point. 
In the sense of the perturbative expansion in \eqref{eq:GellMannLow}, one may first study the limit of interacting fields approaching each other \cite{Zimmermann:1972tv} and second the behavior under changes of the subtraction degree $\de$ \cite{Zimmermann:1972te}. 
Indeed, the BPHZ method in configuration space admits an analogous construction \cite{Pottel:2017cc}, however it was performed only for an arbitrary but fixed order of the perturbative expansion. 
In the following we want to extend the result for two interacting fields $\CA(x_1)\CA(x_2)$ incorporating the full perturbative series and using the reduction formalism \cite{Lehmann:1954rq,Lehmann:1957zz}. \\
We begin with the time-ordered product 
\begin{align}\label{eq:NormalProductStart}
	\VEV{ \CT_R \CA(x_1) \CA(x_2) \CA(y_1)...\CA(y_m)}
\end{align}
with $m$ spectator fields. 
Applying the perturbative argument and expanding in the sense of \eqref{eq:EffectiveGellMannLow}, we obtain external fields $A(x_1)$ and $A(x_2)$ either at the same or at disjoint connected components and furthermore either with or without spectator fields attached. 
In any case, we noticed already above that no renormalization part exists involving two external fields but one type of renormalization part involving one external vertex so that we do not find $A(x_1)$ and $A(x_2)$ in the same renormalization part, but potentially in disjoint renormalization parts simultaneously.
In particular, this means that the coincidence limit $x_j\rightarrow x$ may induce overlap in the forest formula \eqref{eq:ForestFormula}.
Furthermore, the limit may create new renormalization parts.
For this, note that a vertex $A^2(x)$ may be inserted into a two-point function creating a renormalization part with subtraction degree zero according to \eqref{eq:UVDegreeVertex}.
This argument only makes sense in the presence of spectator fields.
Otherwise we can either control
\begin{align}
	\CA(x_1)\CA(x_2) \rightarrow \CA^2(x) 
\end{align} 
by Wick ordering if $A(x_1)$ and $A(x_2)$ belong to the same connected component or divide the contribution out if we manage to relate $N_2[A^2(x)]$ to $N_4[A^2(x)]$ from $\SL_\eff[A,\la](x)$.

In order to quantify our previous analysis, let us consider a graph $\G$, which may or may not consist of several connected components $\G_1,...,\G_c$.
Further, we assign vertices $V_1$ and $V_2$ to the fields $A(x_1)$ and $A(x_2)$, respectively, and denote by $\De$ the graph after the limit $x_j\rightarrow x$, creating the new vertex $V_0$.
We define the operations 
\begin{align}
	\tilde \G = \De \qquad \textrm{\&} \qquad \hat \De = \G 
\end{align}
for completeness.
For the comparison of the renormalized integrands before and after the limit, we assume that $A(x_1)A(x_2)$ is already replaced by its Wick ordered version $:A(x_1)A(x_2):$ for simplicity and we emphasize that due to our definition of subtraction point \eqref{eq:SubtractionPoint}, it is sensible and well-defined to compare the forest formulas of $\G$ and $\De$.
We start with
\begin{align}
	R_\De u_0^\e[\G] = \sum_{F\in\SF_\De} \prod_{\g\in F} (-t(\g)) u_0^\e[\G]
\end{align}
and decompose $\SF_\De$ into the set of all forests $\SF_0$, which contain new renormalization parts, and its complement $\SF_\perp$, for which $\SF_\perp\subset\SF_\G$ holds in general due to possible creation of overlap in the limit, so that
\begin{align}\label{eq:RDeltaDecomposition}
	R_\De u_0^\e[\G] = \underbrace{\sum_{F_\perp \in \SF_\perp} \prod_{\g\in F_\perp} (-t(\g)) u_0^\e[\G]}_{\bydef R_\G u_0^\e[\G] - X_\G u_0^\e[\G]} + \underbrace{\sum_{F_0\in\SF_0} \prod_{\g \in F_0} (-t(\g)) u_0^\e[\G]}_{\bydef X_\De u_0^\e[\G]}.
\end{align}
We observe that for each $F_0\in\SF_0$, there exists a minimal new renormalization part $\ta\subseteq\De$ for which $\si=\hat \ta\subseteq\G$ is not a renormalization part.
Here, minimal refers to the set of new renormalization parts in $F_0$.
Then it follows that each new renormalization part is at least once minimal such that 
\begin{align}\label{eq:NormProdTauOfT}
	X_\De u_0^\e[\G] = \sum_{\ta\in\ST_\De} \sum_{\ol F_\ta \in \ol\SF_\ta} \prod_{\g\in\ol F_\ta} (-t(\g)) (-t(\ta)) \sum_{\ul F_\ta \in \ul\SF_\ta} \prod_{\g'\in \ul F_\ta} (-t(\g')) u_0^\e[\G],
\end{align}
where $\ST_\De$ is the set of all new renormalization parts, $\ol\SF_\ta$ is the set of all $\ta$-superforest, i.e. $\g\supset\ta$ or $\g\cap\ta=\emptyset$ holds for $\g\in\ol F_\ta$, and $\ul\SF_\ta$ is the set of all $\ta$-subforests (or normal $\ta$-forests), i.e. $\g'\subset\ta$ holds for $\g'\in\ul F_\ta$.
Spelling out the Taylor operator $t(\ta)$, we arrive at
\begin{align}\label{eq:NormProdXDeltaGraph}
 	X_\De u_0^\e[\G] = - \sum_{\ta\in\ST_\De} \sum_{|\al|=0}^{\de(\ta)} \frac{1}{\al!} R_{\De/\ta}(D^\al_{\ol V} u_0^\e[\De/\ta]) (x-\ol x_\ta)^\al R_{\ta^\perp} u_0^\e[\si]	
\end{align} 
setting
\begin{align}
	R_{\De/\ta} & = \sum_{\ol F_\ta \in \ol\SF_\ta} \prod_{\g\in\ol F_\ta} (-t(\g)) \\
	R_{\ta^\perp} & = \sum_{\ul F_\ta \in \ul\SF_\ta} \prod_{\g'\in \ul F_\ta} (-t(\g'))
\end{align}
and using
\begin{align}
	u_0^\e[\G\lineco\si] = u_0^\e[\De\lineco\ta],
\end{align}
where vertex $\ol V$ results from the contraction of $\ta$ to a point in $\De$.
If additionally
\begin{align}\label{eq:OlapCreationContra}
	R_{\ta^\perp} u_0^\e[\si] = R_{\si^\perp} u_0^\e[\si]
\end{align}
would hold, then $X_\G  u_0^\e[\G]\equiv 0$ would follow.
However, \eqref{eq:OlapCreationContra} cannot be true in general due to overlap creation in the coincidence limit. 
Specifically, consider all pairs $(\z_1,\z_2)$ of mutually disjoint renormalization parts $\z_j\subset\G$ with $V_1\in V(\z_1)$ and $V_2\in V(\z_2)$ and subsume those pairs in a set $\SZ$.
Then we get
\begin{align}
	X_\G u_0^\e[\G] = \sum_{(\z_1,\z_2)\in\SZ} \sum_{\ol F_\z \in \ol\SF_\z} \prod_{\g\in\ol F_\z} (-t(\g)) (-t(\z_1))(-t(\z_2)) \sum_{\ul F_\z \in \ul\SF_\z} \prod_{\g'\in \ul F_\z} (-t(\g')) u_0^\e[\G],
\end{align}
which turns into
\begin{multline}
	X_\G u_0^\e[\G] = \sum_{(\z_1,\z_2)\in\SZ} \sum_{|\al_1|=0}^{\de(\z_1)} \sum_{|\al_2|=0}^{\de(\z_2)} \frac{1}{\al_1! \al_2!} R_{\G/\z}(D^{\al_1}_{\ol V_1} D^{\al_2}_{\ol V_2} u_0^\e[\G/\z]) \times \\
	\times (x - \ol x_{\z_1})^{\al_1} R_{\z_1^\perp} u_0^\e[\z_1] (x-\ol x_{\z_2})^{\al_2} R_{\z_2^\perp} u_0^\e[\z_2].
\end{multline}
Next, recall from the discussion at the beginning of this section that all new renormalization parts $\ta$ as well as all overlap creating renormalization parts $\z_j$ have UV-degree of divergence zero naively so that
\begin{multline}
	R_\G u_0^\e[\G] - R_\De u_0^\e[\G] = \sum_{\ta\in\ST_\De} R_{\De/\ta} u_0^\e[\De/\ta] R_{\ta^\perp} u_0^\e[\si] \\ + \sum_{(\z_1,\z_2)\in\SZ} R_{\G/\z}u_0^\e[\G/\z] R_{\z_1^\perp} u_0^\e[\z_1] R_{\z_2^\perp} u_0^\e[\z_2].
\end{multline}
Only the summation over all graphs $\G$ is left and, again, we refer to the discussion of renormalization parts contributing to $X_\De u_0^\e[\G]$ and $X_\G u_0^\e[\G]$. 
Omitting integrations, we obtain for the overlap creation
\begin{align}
	\sum R_{\z_j^\perp}u_0^\e[\si] \simeq \underbrace{\VEV{\CT_R A(x_j) \int A^3(z_j) dz_j  \exp\left\{ i\int \SL_\eff[A,\la](y)dy \right\} }^\mathrm{conn}}_{\bydef \SCL_1(x_j)}
\end{align}
and for the new renormalization parts either
\begin{align}
	\sum R_{\ta^\perp} u_0^\e[\si] \simeq \underbrace{\VEV{\CT_R :A(x_1)A(x_2): \int A^2(z) dz \exp\left\{ i\int \SL_\eff[A,\la](y)dy \right\}}^\mathrm{conn}}_{\bydef \SCL_2(x_1,x_2)}
\end{align}
if it is tadpole-like
\begin{multline}\label{eq:CoeffFuncSingle}
	\sum R_{\ta^\perp} u_0^\e[\si] \simeq \underbrace{\VEV{\CT_R :A(x_1)A(x_2): \int A^3(z_1) dz_1 \int A^3(z_2) dz_2 \exp\left\{ i\int \SL_\eff[A,\la](y)dy \right\}}^\mathrm{conn}}_{\bydef \SCL_3(x_1,x_2)}
\end{multline}
if it stems from a single connected component or
\begin{multline}\label{eq:CoeffFuncMultiple}
	\sum R_{\ta^\perp} u_0^\e[\si] \simeq \VEV{\CT_R A(x_1) \int A^3(z_1) dz_1 \exp\left\{ i\int \SL_\eff[A,\la](y)dy \right\}}^\mathrm{conn} \times \\
	\times \VEV{\CT_R A(x_2) \int A^3(z_2) dz_2 \exp\left\{ i\int \SL_\eff[A,\la](y)dy \right\}}^\mathrm{conn}
\end{multline}
if it stems from multiple connected components.
The corresponding contributions with contracted renormalization part(s) are given by (omitting non-participating connected components)
\begin{align}\label{eq:NewRPartsTProduct}
	\VEV{\CT_R A^2(\ol x) A(y_{i_1}) ... A(y_{i_c}) \exp\left\{ i\int \SL_\eff[A,\la](y)dy \right\}}^\mathrm{conn}
\end{align}
for new renormalization parts and by either
\begin{multline}\label{eq:OlapCreationMultiple}
	\VEV{\CT_R A(x_1) A(y_{j_1}) ... A(y_{j_a}) \exp\left\{ i\int \SL_\eff[A,\la](y)dy \right\}}^\mathrm{conn} \times \\
	\times \VEV{\CT_R A(x_2) A(y_{j_{a+1}}) ... A(y_{j_b}) \exp\left\{ i\int \SL_\eff[A,\la](y)dy \right\}}^\mathrm{conn}
\end{multline}
for multiple connected components or
\begin{align}\label{eq:OlapCreationSingle}
	\VEV{\CT_R :A(x_1)A(x_2): A(y_{k_1})...A(y_{k_b}) \exp\left\{ i\int \SL_\eff[A,\la](y)dy \right\}}^\mathrm{conn}
\end{align}
for a single connected component in the case of overlap creation.
We observe that \eqref{eq:OlapCreationMultiple} and \eqref{eq:OlapCreationSingle} (analogously for \eqref{eq:CoeffFuncSingle} and \eqref{eq:CoeffFuncMultiple}) may be rewritten as
\begin{align}
	\frac{\VEV{\CT_R :A(x_1)A(x_2): A(y_1)...A(y_m)  \exp\left\{ i\int \SL_\eff[A,\la](y)dy \right\}}}{\VEV{\CT_R \exp\left\{ i\int \SL_\eff[A,\la](y)dy \right\}}}
\end{align}
with $m\geq b$, which coincides with \eqref{eq:NormalProductStart} up to Wick ordering that we discussed above. 
However, $\SCL_1(x_j)$ is a logarithmically divergent contribution adjacent to $A(x_j)$ and, as such, vanishes identically after application of the $R$-operation.
Hence we obtain
\begin{multline}
	\VEV{\CT_R N_2[\CA(x_1) \CA(x_2)] \CA(y_1)... \CA(y_m)}
	= \VEV{\CT_R :\CA(x_1)\CA(x_2): \CA(y_1)... \CA(y_m)} \\
	- (\SCL_2(x_1,x_2) + \SCL_3(x_1,x_2)) \VEV{\CT_R N_2[\CA^2(\ol x)] \CA(y_1)... \CA(y_m)}
\end{multline}
such that $N_2[\CA(x_1)\CA(x_2)]$ admits the limit $\x_j\rightarrow x$ \cite{Pottel:2017cc}.
In order to conclude, we spell out the Wick ordering explicitly
\begin{multline}
	\VEV{\CT_R :\CA(x_1)\CA(x_2): \CA(y_1)... \CA(y_m)} \\
	= \VEV{\CT_R \CA(x_1)\CA(x_2) \CA(y_1)... \CA(y_m)} - \underbrace{\VEV{\CT_R \CA(x_1)\CA(x_2)}}_{\bydef \SCL_0(x_1,x_2)} \VEV{\CT_R \CA(y_1)... \CA(y_m)}
\end{multline}
and, exploiting the translation invariance, find after applying the reduction formalism that
\begin{align}
	\CT_R \CA(\ol x + \x)\CA(\ol x-\x) = \SCL_0(\x) \II + (\SCL_2(\x) + \SCL_3(\x)) N_2[\CA^2(\ol x)] +r(\ol x, \x),
\end{align}
where $\ol x = \frac{1}{2}(x_1+x_2)$, $\x = \frac{1}{2}(x_1-x_2)$ and for $\x\rightarrow 0$
\begin{align}
	\SCL_0(\x) & \simeq \CO(|\x|^{-2}) \\
	\SCL_2(\x) & \simeq \CO(\log(\x)) \\
	\SCL_3(\x) & \simeq \CO(\log(\x)) \\
	r(\ol x,\x) & \rightarrow 0
\end{align}
holds, which is in accordance with the findings in \cite{Zimmermann:1972tv}.

\section{Zimmermann Identity}\label{se:ZI}

Next, we want to discuss the relation of the composite operator $N_2[\CA^2(x)]$ to oversubtraction degree, which is particularly interesting in view of interaction vertices $N_4[A^2(z)]$ in $\SL_\eff[A,\la](z)$.
For simplicity and in regard to the discussion of possible renormalization parts (recall \eqref{eq:UVDegreeVertex}), we restrict ourselves to one insertion
\begin{align}
	\VEV{\CT \CA(x_1)...\CA(x_n) N_\de[\CA^2(x)]},\qquad \de\in\{ 2, 4\}
\end{align}
with varying subtraction degree.
Without loss of generality, we consider only connected components $\De$ in the perturbative expansion and assign the vertex $V_0$ to $N_\de[A^2(x)]$.
Again, we split the forest formula for a graph $\De$ into the set $\SF_0$ of all forests $F_0$, which contain a renormalization part with $V_0$ in its vertex set, and the complement $\SF_\perp$, i.e.
\begin{align}
	R_\de u_0^\e[\De] = \underbrace{\sum_{F_0\in\SF_0} \prod_{\g\in F_0} (-t_{(\de)}(\g)) u_0^\e[\De]}_{\bydef X u_0^\e[\De]} + \sum_{F_\perp\in\SF_\perp} \prod_{\g\in F_\perp} (-t_{(\de)}(\g)) u_0^\e[\De].
\end{align}
Note that this splitting does generally not coincide with the prescription in \eqref{eq:RDeltaDecomposition} and that we do not have to be concerned with overlap creation in particular.
Furthermore, we have $t_{(\de)}(\g)=t(\g)$ for $\g\in F_\perp$ so that only $Xu_0^\e[\De]$ is affected by changes in the subtraction degree of $N_\de[A^2(x)]$.
Beginning with $N_4[A^2(x)]$, there exists a minimal renormalization part $\ta\in F_0$ with $V_0\in V(\ta)$ in each $F_0$, for which we rewrite
\begin{align}\label{eq:TaylorFourToTaylorTwo}
 	t_{(4)}(\ta) = t_{(2)}(\ta) + (t_{(4)}(\ta)-t_{(2)}(\ta))	
\end{align} 
and expand the right-had side of \eqref{eq:TaylorFourToTaylorTwo} in $Xu_0^\e[\G]$.
Each renormalization part $\ta$ with $V_0\in V(\ta)$ is at least once minimal with respect to the Taylor difference operator $(t_{(4)}(\ta)-t_{(2)}(\ta))$ such that all supergraphs $\g\supset\ta$ are subtracted by prescription $\de=4$ and all subgraphs $\g'\subset\ta$ are subtracted by prescription $\de=2$.
In analogy to \eqref{eq:NormProdTauOfT}, it is straightforward to show
\begin{align}
	Xu_0^\e[\De] = \sum_{\ta\in\ST} \sum_{\ol F_\ta \in \ol\SF_\ta} \prod_{\g\in\ol F_\ta} (-t(\g)) (-(t_{(4)}(\ta)-t_{(2)}(\ta))) \sum_{\ul F_\ta \in \ul\SF_\ta} \prod_{\g'\in \ul F_\ta} (-t(\g')) u_0^\e[\De],
\end{align}
where the Taylor difference operator becomes either $(t^2(\ta)-t^0(\ta))$ in case of two external legs, indicated by $\ST_2$, or simplifies to $t^0(\ta)$ in case of four external legs, indicated by $\ST_4$, so that
\begin{align}
	Xu_0^\e[\De] = & \sum_{\ta\in\ST_2} \sum_{|\al|=1}^2 \frac{1}{\al!} R_{\De/\ta} D^\al_{\ol V_1} u_0^\e[\De/\ta] (x-\ol x_\ta)^\al R_{\ta^\perp} u_0^\e[\ta] \\
	& + \sum_{\ta\in\ST_4} R_{\De/\ta} D^\al_{\ol V_2} u_0^\e[\De/\ta] R_{\ta^\perp} u_0^\e[\ta],
\end{align} 
where $\ol V_1$ has valency $2$ and $\ol V_2$ has valency $4$.
It remains to perform the summation over all graphs $\De$, where the vertices $\ol V_1$ and $\ol V_2$ still have to be compatible with the $A^4$-model.
Similar to the treatment of the coincidence limit, we may split the sum over all graphs $\De$ and all renormalization parts $\ta\subseteq\De$ into two independent sums over graphs containing either a vertex $\ol V_1$ or a vertex $\ol V_2$, all of them being subtracted by the prescription $\de=4$, and over graphs, which are the corresponding renormalization part to either $\ol V_1$ or $\ol V_2$, all of them being subtracted by the prescription $\de=2$. 
The admissible vertices are given by
\begin{align}
	\SG_1^{\m\n} N_4[A(x)\pa_\m \pa_\n A(x)] ,\quad  \SG_2^{\m\n} N_4[\pa_\m A(x)\pa_\n A(x)] &\mbox{ for $|\al|=2$} \\
	\SG_3^\m N_4[A(x) \pa_\m A(x)] & \mbox{ for $|\al|=1$} \\
	\SG_4 N_4[A^4(x)] & \mbox{ for $|\al|=0$}
\end{align}  
with
\begin{align}
	\SG_1^{\m\n} & = \VEV{\CT_R N_2[A^2(x)] \int A^3(y_1) \int A^3(y_2) (x-\ol x)^{\m_j} (x-\ol x)^{\n_j} dy_2 dy_1 \exp\left\{ i\int \SL_\eff[A,\la](y)dy \right\}}^\mathrm{conn}\\
	\SG_2^{\m\n} & = \VEV{\CT_R N_2[A^2(x)] \int A^3(y_1) \int A^3(y_2) (x-\ol x)^{\m_j} (x-\ol x)^{\n_i} dy_2 dy_1 \exp\left\{ i\int \SL_\eff[A,\la](y)dy \right\}}^\mathrm{conn}\\
	\SG_3^\m & = \VEV{\CT_R  N_2[A^2(x)] \int A^3(y_1) \int A^3(y_2) (x-\ol x)^{\m_j} dy_2 dy_1 \exp\left\{ i\int \SL_\eff[A,\la](y)dy \right\}}^\mathrm{conn} \\
	\SG_4 & = \VEV{\CT_R N_2[A^2(x)] \int A^2(y_1) dy_1 \int A^2(y_2) dy_2 \exp\left\{ i\int \SL_\eff[A,\la](y)dy \right\}}^\mathrm{conn},
\end{align}
where the moments are summed over all $i,j$, $i\neq j$, running through all interaction vertices and $V_0$. 
We remark that all graphs $\De$ are already locally integrable choosing the prescription with $\de=2$ so that each $\SG_j^\bullet$ is finite at each order of the expansion in particular.
Moreover, they are constant coefficients due to translation invariance, which reduces the number of possible vertices to
\begin{align}
	\SG_1 N_4[A(x) \Box A(x)] ,\quad  \SG_2 N_4[\pa_\m A(x)\pa^\m A(x)] ,\quad \SG_4 N_4[A^4(x)].
\end{align}
Applying the reduction formalism, we arrive at the Zimmermann identity
\begin{align}\label{eq:ZI}
	N_2[\CA^2(x)] - N_4[\CA^2(x)] 
	= \SG_1 N_4[\CA(x) \Box \CA(x)] + \SG_2 N_4[\pa_\m \CA(x)\pa^\m \CA(x)] + \SG_4 N_4[\CA^4(x)].
\end{align}

\section{Conclusion}
\label{se:Conc}

In Section \ref{se:Pres}, we motivated our work by the analysis of a perturbative treatment of the equation of motion \eqref{eq:ClassicalNonLinEoM}.
We want to return to the equation of motion and study the linear equation
\begin{align}\label{eq:FreeKleinGordon}
	(\Box + m^2) A(x) = 0
\end{align}
as insertion 
\begin{align}
	N_3\left[ \frac{\de\SL_0[A]}{\de(\pa_\m A)}(x) - \frac{\de\SL_0[A]}{\de A} (x) \right]
\end{align}
into some $n$-point function
\begin{align}
	\VEV{\CT_R \CA(x_1)...\CA(x_n)}.
\end{align}
Note that $A(x)$ does not fulfill \eqref{eq:FreeKleinGordon} in a time-ordered product. 
Instead, every contraction of $A(x)$ with another field leads to a propagator $G_c(x-\bullet)$ such that 
\begin{align}
	(\Box+m^2) G_c(x-\bullet) = -i \de(x-\bullet)
\end{align}
holds for the insertion.
Evaluating the integration over the free variable (in case it is not another external vertex), leaves us with a fused vertex at locus $x$.
Let us denote by $\G$ and $\De$ the graph before and after the fusion process, respectively.
We observe that in this process neither overlap creation nor Zimmermann identity terms can appear and the forest formulas remain unchanged.
Then we obtain
\begin{align}
	\VEV{\CT_R N_3[(\Box+m^2)\CA(x)] \CA(x_1)...\CA(x_n)} = & \sum_{j=1}^n \de(x-x_j) \VEV{\CT_R \CA(x_1)...\CA(\check x_j)...\CA(x_n)} \\ 
	& - \VEV{\CT_R N_3[a\Box \CA(x)]\CA(x_1)...\CA(x_n)} \\
	& + \VEV{\CT_R N_3[b \CA(x)] \CA(x_1)...\CA(x_n)} \\
	& + \VEV{\CT_R \frac{1}{3!} N_3[c \CA^3(x)] \CA(x_1)...\CA(x_n)}, 
\end{align}
which becomes for connected graphs
\begin{multline}
	\VEV{\CT_R N_3[((1+a)\Box+m^2_\mathrm{phys})\CA(x)] \CA(x_1)...\CA(x_n)}^\mathrm{conn}\\
	= \VEV{\CT_R \frac{1}{3!} N_3[c \CA^3(x)] \CA(x_1)...\CA(x_n)}^\mathrm{conn},
\end{multline}
where $m^2_\mathrm{phys} = m^2 - b$. 
Furthermore, using the reduction formalism, the equation of motion becomes
\begin{align}
	((1+a)\Box + m^2_\mathrm{phys})\CA(x) = \frac{c}{3!} \CA^3(x).
\end{align}
Instead of considering the Klein-Gordon operator, it is possible to study differentiation with respect to parameters of the theory or field operators as well. 
Recall that we defined the perturbative expansion via the effective Lagrangian $\SL_\eff[A,\la]$, i.e. we introduced interactions
\begin{align}
	\De_1 &\bydef \int N_4[\pa_\m A(y) \pa_\m A(x)] dy \\
	\De_2 &\bydef \int N_4[A^2(y)] dy \\
	\De_4 &\bydef \int N_4[A^4(y)] dy 
\end{align}
together with their corresponding parameter $a$, $b$ and $c$.
Then we differentiate with respect to mass
\begin{align}
	m^2 \frac{\pa}{\pa m^2} \exp\left\{ i\int \SL_\eff[A,\la](y)dy \right\} = (m^2 + b) \De_2 \exp\left\{ i\int \SL_\eff[A,\la](y)dy \right\},
\end{align}
coupling
\begin{align}
	\frac{\pa}{\pa \la} \exp\left\{ i\int \SL_\eff[A,\la](y)dy \right\}
	= \left( \frac{\pa a}{\pa \la} \De_1 + \frac{\pa b}{\pa \la} \De_2 + \frac{\pa c}{\pa \la} \De_4 \right) \exp\left\{ i\int \SL_\eff[A,\la](y)dy \right\},
\end{align}
and the field 
\begin{align}\label{eq:NumberOperator}
	A \frac{\de}{\de A} \exp\left\{ i\int \SL_\eff[A,\la](y)dy \right\} 
	= \left(2 a \De_1 + 2 b \De_2 + 4 c \De_4 \right) \exp\left\{ i\int \SL_\eff[A,\la](y)dy \right\},
\end{align}
which admits the derivation of the Callan-Symanzik or renormalization group equation \cite{Lowenstein:1971jk} using the Zimmermann identity \eqref{eq:ZI}.
Alternatively, one can study the response of $n$-point functions to variations of the field $\de^Q A$, by some transformation $Q$, via Ward operators \cite{Piguet:1980nr,Kraus:1991cq,Kraus:1992ru,Pottel:2010hp}
\begin{align}
	W^Q \bydef \int dz \de^Q A(z) \frac{\de}{\de A(z)}.
\end{align}
The study of anomalies for theories with massless propagators should be of particular interest, since, differently from the momentum space prescription, the introduction of an auxiliary mass term \cite{Lowenstein:1974qt} is not necessary.
In fact, it is not even admissible in view of the extension problem \cite{Pottel:2017bb}.

\subsection*{Acknowledgments}

The author would like to thank Klaus Sibold for numerous discussions. The financial support by the Max Planck Institute for Mathematics in the Sciences and its International Max Planck Research
School (IMPRS) ``Mathematics in the Sciences'' is gratefully acknowledged. 

%% Bib
\bibliographystyle{halpha} 
\bibliography{bphzl}

\begin{thebibliography}{Low71b}

\bibitem[BF00]{Brunetti:1999jn}
R.~Brunetti and K.~Fredenhagen.
\newblock {Microlocal analysis and interacting quantum field theories:
  Renormalization on physical backgrounds}.
\newblock {\em Commun.Math.Phys.}, 208:623--661, 2000, math-ph/9903028.

\bibitem[BGP07]{Bar:2007zz}
C.~Bar, N.~Ginoux, and F.~Pfaffle.
\newblock {\em {Wave equations on Lorenzian manifolds and quantization}}.
\newblock 2007.

\bibitem[BP57]{Bogoliubov:1957gp}
N.~N. Bogoliubov and O.~S. Parasiuk.
\newblock {On the Multiplication of the causal function in the quantum theory
  of fields}.
\newblock {\em Acta Math.}, 97:227--266, 1957.

\bibitem[BRS76]{Becchi:1975nq}
C.~Becchi, A.~Rouet, and R.~Stora.
\newblock {Renormalization of Gauge Theories}.
\newblock {\em Annals Phys.}, 98:287--321, 1976.

\bibitem[BS59]{Bogolyubov:1980nc}
N.N. Bogolyubov and D.V. Shirkov.
\newblock {Introduction to the Theory of Quantized Fields}.
\newblock {\em Intersci.Monogr.Phys.Astron.}, 3:1--720, 1959.

\bibitem[Col86]{Collins:1984xc}
J.~C. Collins.
\newblock {\em {Renormalization}}, volume~26 of {\em Cambridge Monographs on
  Mathematical Physics}.
\newblock Cambridge University Press, Cambridge, 1986.

\bibitem[Hep66]{Hepp:1966eg}
K.~Hepp.
\newblock {Proof of the Bogolyubov-Parasiuk theorem on renormalization}.
\newblock {\em Commun.Math.Phys.}, 2:301--326, 1966.

\bibitem[H{\"o}r90]{hormander1990analysis}
L.~H{\"o}rmander.
\newblock {\em The analysis of linear partial differential operators:
  Distribution theory and Fourier analysis}.
\newblock Springer Study Edition. Springer-Verlag, 1990.

\bibitem[KS92]{Kraus:1991cq}
E.~Kraus and K.~Sibold.
\newblock {Conformal transformation properties of the energy momentum tensor in
  four-dimensions}.
\newblock {\em Nucl. Phys.}, B372:113--144, 1992.

\bibitem[KS93]{Kraus:1992ru}
E.~Kraus and K.~Sibold.
\newblock {Local couplings, double insertions and the Weyl consistency
  condition}.
\newblock {\em Nucl. Phys.}, B398:125--154, 1993.

\bibitem[Low71a]{Lowenstein:1971jk}
J.~H. Lowenstein.
\newblock {Differential vertex operations in Lagrangian field theory}.
\newblock {\em Commun. Math. Phys.}, 24:1--21, 1971.

\bibitem[Low71b]{Lowenstein:1971vf}
J.~H. Lowenstein.
\newblock {Normal product quantization of currents in Lagrangian field theory}.
\newblock {\em Phys. Rev.}, D4:2281--2290, 1971.

\bibitem[Low76]{Lowenstein:1975ps}
J.~H. Lowenstein.
\newblock {Convergence Theorems for Renormalized Feynman Integrals with
  Zero-Mass Propagators}.
\newblock {\em Commun. Math. Phys.}, 47:53--68, 1976.

\bibitem[LS76]{Lowenstein:1975ku}
J.~H. Lowenstein and E.~R. Speer.
\newblock {Distributional Limits of Renormalized Feynman Integrals with
  Zero-Mass Denominators}.
\newblock {\em Commun. Math. Phys.}, 47:43--51, 1976.

\bibitem[LSZ55]{Lehmann:1954rq}
H.~Lehmann, K.~Symanzik, and W.~Zimmermann.
\newblock {On the formulation of quantized field theories}.
\newblock {\em Nuovo Cim.}, 1:205--225, 1955.

\bibitem[LSZ57]{Lehmann:1957zz}
H.~Lehmann, K.~Symanzik, and W.~Zimmermann.
\newblock {On the formulation of quantized field theories. II}.
\newblock {\em Nuovo Cim.}, 6:319--333, 1957.

\bibitem[LZ75a]{Lowenstein:1974qt}
J.~H. Lowenstein and W.~Zimmermann.
\newblock {On the Formulation of Theories with Zero Mass Propagators}.
\newblock {\em Nucl. Phys.}, B86:77--103, 1975.

\bibitem[LZ75b]{Lowenstein:1975rg}
J.~H. Lowenstein and W.~Zimmermann.
\newblock {The Power Counting Theorem for Feynman Integrals with Massless
  Propagators}.
\newblock {\em Commun. Math. Phys.}, 44:73--86, 1975.

\bibitem[Pot17a]{Pottel:2017aa}
S.~Pottel.
\newblock {A BPHZ Theorem in Configuration Space}.
\newblock 2017, 1706.06762.

\bibitem[Pot17b]{Pottel:2017bb}
S.~Pottel.
\newblock {Configuration Space BPHZ Renormalization on Analytic Spacetimes}.
\newblock 2017, 1708.04112.

\bibitem[Pot17c]{Pottel:2017cc}
S.~Pottel.
\newblock {Normal Products and Zimmermann Identities in Configuration Space
  BPHZ Renormalization}.
\newblock 2017, 1708.04115.

\bibitem[PR81]{Piguet:1980nr}
O.~Piguet and A.~Rouet.
\newblock {Symmetries in Perturbative Quantum Field Theory}.
\newblock {\em Phys. Rept.}, 76:1, 1981.

\bibitem[PS10]{Pottel:2010hp}
S.~Pottel and K.~Sibold.
\newblock {Conformal Transformations of the S-Matrix: $\beta$-Function
  Identifies Change of Spacetime}.
\newblock {\em Phys. Rev.}, D82:025001, 2010, 1004.3180.

\bibitem[Ste00]{Steinmann:2000nr}
O.~Steinmann.
\newblock {\em {Perturbative quantum electrodynamics and axiomatic field
  theory}}.
\newblock 2000.

\bibitem[Tyu75]{Tyutin:1975qk}
I.~V. Tyutin.
\newblock {Gauge Invariance in Field Theory and Statistical Physics in Operator
  Formalism}.
\newblock 1975, 0812.0580.

\bibitem[WZ72]{Wilson:1972ee}
K.~G. Wilson and W.~Zimmermann.
\newblock {Operator product expansions and composite field operators in the
  general framework of quantum field theory}.
\newblock {\em Commun. Math. Phys.}, 24:87--106, 1972.

\bibitem[Zim68]{Zimmermann:1968mu}
W.~Zimmermann.
\newblock {The power counting theorem for minkowski metric}.
\newblock {\em Commun.Math.Phys.}, 11:1--8, 1968.

\bibitem[Zim69]{Zimmermann:1969jj}
W.~Zimmermann.
\newblock {Convergence of Bogolyubov's method of renormalization in momentum
  space}.
\newblock {\em Commun. Math. Phys.}, 15:208--234, 1969.

\bibitem[Zim73a]{Zimmermann:1972te}
W.~Zimmermann.
\newblock {Composite operators in the perturbation theory of renormalizable
  interactions}.
\newblock {\em Annals Phys.}, 77:536--569, 1973.

\bibitem[Zim73b]{Zimmermann:1972tv}
W.~Zimmermann.
\newblock {Normal products and the short distance expansion in the perturbation
  theory of renormalizable interactions}.
\newblock {\em Annals Phys.}, 77:570--601, 1973.

\end{thebibliography}

\end{document}